\documentclass[aps,prl,twocolumn,groupedaddress]{revtex4}
\usepackage{graphicx}% Include figure files
\usepackage{dcolumn}% Align table columns on decimal point
\usepackage{bm}% bold math
\usepackage{amsmath}
\usepackage{xspace}
\usepackage{longtable}
\newcommand{\ttbar}{t\bar{t}}
\newcommand{\tbbar}{t\bar{b}}
\newcommand{\bbbar}{b\bar{b}}
\newcommand{\ccbar}{c\bar{c}}
\newcommand{\pbarp}{{\bar p}p}

\newcommand{\roots}{{\sqrt s}}
 
\newcommand{\Et}{E_T}
\newcommand{\Pt}{p_T}

\newcommand{\gt}{>}
\newcommand{\lt}{<}

\newcommand{\mett}{{\not\!\!E}_{T}}
\newcommand{\Dzero}{D\O~}
\newcommand{\MeV}{\ensuremath{\text{\ Me\kern -0.1em V}}\xspace}

\newcommand{\MeVc}{\ensuremath{\text{\ Me\kern -0.1em V\kern -0.1em 
/\mathit{c}}}\xspace}

\newcommand{\MeVcsq}{\ensuremath{\text{\ Me\kern -0.1em V\kern -0.1em 
/\mathit{c}^2}}\xspace}

\newcommand{\GeV}{\ensuremath{\text{\ Ge\kern -0.1em V}}\xspace}

\newcommand{\GeVc}{\ensuremath{\text{\ Ge\kern -0.1em V\kern -0.1em 
/\mathit{c}}}\xspace}

\newcommand{\GeVcsq}{\ensuremath{\text{\ Ge\kern -0.1em V\kern -0.1em 
/\mathit{c}^2}}\xspace}

\newcommand{\TeV}{\ensuremath{\text{\ Te\kern -0.1em V}}\xspace}

\newcommand{\nsGeV}{\ensuremath{\text{Ge\kern -0.1em V}}\xspace}

\newcommand{\nsGeVc}{\ensuremath{\text{Ge\kern -0.1em V\kern -0.1em /\mathit{c}}}\xspace}

\newcommand{\ellnu}{\ell\nu}

\newcommand{\ellpellm}{\ell^+\ell^-}

\newcommand{\Z}{\ensuremath{Z^0}~}

\newcommand{\Zg}{\ensuremath{Z^0 / \gamma^*}~}
\newcommand{\Wp}{\ensuremath{W^{+}}}
\newcommand{\WpNjets}{\ensuremath{W^{+}+N~ \text{jets}}~}

\newcommand{\WpbbNjets}{\ensuremath{W^{+}+\bbbar+N~\text{jets}}~}
\newcommand{\WpenuNjets}{\ensuremath{W^{+}(\rightarrow e^+\nu) ~+N~
\text{jets}}~}
\newcommand{\Wpenujet}{\ensuremath{W^{+}(\rightarrow e^+\nu) ~+1~
\text{jet}}~}
\newcommand{\WNjets}{\ensuremath{W+N~ \text{jets}}~}
\newcommand{\ZNjets}{\ensuremath{\Zg +N~ \text{jets}}~}

\newcommand{\ZbbNjets}{\ensuremath{\Zg+\bbbar+N~\text{jets}}~}
\newcommand{\ZeeNjets}{\ensuremath{\Zg(\rightarrow e^+e^-) +N~ \text{jets}}~}
\newcommand{\Zeejet}{\ensuremath{\Zg(\rightarrow e^+e^-) +1~ \text{jet}}~}
\newcommand{\WNtoZN}{\ensuremath{\sigma(W+N \text{jets})/\sigma(\Zg+N \text{jets})}~}
\newcommand{\WpNtoZN}{\ensuremath{\sigma(W^{+}+N
\text{jets})/\sigma(\Zg+N \text{jets})}~}
\newcommand{\WpTWOtoZTWO}{\ensuremath{\sigma(W^{+}+2
\text{jets})/\sigma(\Zg+2 \text{jets})}~}
\newcommand{\WNtoWNp}{\ensuremath{\sigma(W+N \text{jets})/\sigma(W+N+1
\text{jets})}~}
\newcommand{\WpNtoWpNp}{\ensuremath{\sigma(W^{+}+N
\text{jets})/\sigma(W^{+}+N+1 \text{jets})}~}
\newcommand{\ZNtoZNp}{\ensuremath{\sigma(\Zg+N\text{jets})/\sigma(\Zg+N+1
\text{jets})}~}
\newcommand{\sigmaWN}{\ensuremath{W^+_N(Q^2)}~}
\newcommand{\sigmaWNm}{\ensuremath{W^+_N(M^2)}~}
\newcommand{\sigmaWNpm}{\ensuremath{W^+_N(P_t^2+M^2)}~}
\newcommand{\sigmaZN}{\ensuremath{Z_N(Q^2)}~}
\newcommand{\sigmaZNm}{\ensuremath{Z_N(M^2)}~}
\newcommand{\sigmaZNpm}{\ensuremath{Z_N(P_t^2+M^2)}~}

\newcommand{\Qsq}{\ensuremath{Q^2}}

\newcommand{\Qonesq}{\ensuremath{{{\text{Q}}_1}^2}}

\newcommand{\Qtwosq}{\ensuremath{{{\text{Q}}_2}^2}}

\newcommand{\QsqeqMP}{\ensuremath{\text{Q}^2 = {\text{M}_{\text{V}}}^2 +{\text{P}_{\text{T,V}}}^2}}

\newcommand{\QsqeqM}{\ensuremath{\text{Q}^2 = {\text{M}_{\text{V}}}^2}}

\newcommand{\RX}{\ensuremath{\text{R}^{+}}_\text{X}}

\newcommand{\WX}{\ensuremath{\text{W}^{+}}_\text{X}}

\newcommand{\ZX}{\ensuremath{\text{Z}}_\text{X}}

\newcommand{\RL}{\ensuremath{\text{R}^{+}}_{\text{L}}}

\newcommand{\WL}{\ensuremath{\text{W}^{+}}_{\text{L}}}

\newcommand{\ZL}{\ensuremath{\text{Z}}_{\text{L}}}

\newcommand{\Wenu}{W \rightarrow e \nu}

\newcommand{\Wlnu}{W \rightarrow \ell \nu}

\newcommand{\Zll}{\Z \rightarrow \ell^+ \ell^-}

\begin{document}

\preprint{EFI-03-08}

%\preprint{}

\title{The Ratio of \WNjets To \ZNjets Versus $N$ As a
Precision Test of the Standard Model}



\author{Erin Abouzaid}

\author{Henry Frisch}
\affiliation{Enrico Fermi Institute, University of Chicago}

%\thanks{}


\date{\today}

\begin{abstract}

We suggest replacing measurements of the individual cross-sections for the
production of $W+$ N jets and $Z+$ N jets in searches for new
high-energy phenomena at hadron colliders by the precision measurement
of the ratios (W+0~jet)/(Z+0~jet), (W+1~jet)/(Z+1~jet),
(W+2~jets)/(Z+2~jets),... (W+N~jets)/(Z+N~jets), with N as large as 6
(the number of jets in $\ttbar H$).
These ratios can also be formed for the case where one or more of the
jets is tagged as a $b$ or $c$ quark.  Existing measurements of the
individual cross sections for $\Wenu+$ N jets at the Tevatron have systematic
uncertainties that grow rapidly with N, being dominated by
uncertainties in the identification of jets and the jet energy
scale. These systematics, and also those associated with the
luminosity, parton distribution functions (PDF's), detector acceptance
and efficiencies, and systematics of jet finding and b-tagging, are
expected to substantially cancel in calculating the ratio of W to Z
production in each N-jet channel, allowing a greater sensitivity to
new contributions in these channels in Run II at the Tevatron and at
the LHC. 

\end{abstract}

%\keywords{}

\maketitle


\section{\label{section:introduction}Introduction}

The signatures of the leptonic decays of the 
heavy gauge bosons W or \Z accompanied by jets, $\Wlnu+$ jets and
$\Zll+$ jets,
are among the preeminent-eminent search channels in very high energy particle
collisions for `new physics', i.e. interactions or particles that are
not part of the Standard Model~\cite{SM1,SM2,SM3,SM4}.  Many
extensions of the SM predict new particles which have electroweak
(EWK) couplings and decay into the SM gauge bosons W, \Z, and
$\gamma$, accompanied by jets. For example, searches have been made in
the W or \Z +jets channels for supersymmetric
particles~\cite{stop,sbottom}, technicolored
hadrons~\cite{technihadrons}, heavy $W^\prime$ and $Z^\prime$
bosons~\cite{Wprime_D0,Wprime_CDF_1,Wprime_CDF_2} that might arise in
extended gauge groups or from excitations in extra spatial dimensions,
charged Higgs bosons~\cite{chargedHiggs_D0,chargedHiggs_CDF}, and
leptoquarks
~\cite{leptoquarks1_D0,leptoquarks1_CDF,leptoquarks2_D0,leptoquarks2_CDF},
among others. More generally, any production of new heavy particles
with quantum numbers conserved by the strong interaction and EWK
couplings is likely to contribute to signatures with one or more EWK
gauge bosons; additional jets will always be present at some level
from initial-state radiation, and may also be created in cascade
decays of new heavy particles or from the decay of associated heavy
particles.

Within the SM, the top quark was discovered and its mass measured in
the W+3/4 jets channel in which at least one jet was identified as a
b-quark
~\cite{top_CDF_ev_PRL,top_CDF_ev_PRD,top_CDF_observation,top_D0_observation}.
The W+ 2 jets channel with b-quark identification has been used to
search for the Higgs boson~\cite{SMHiggs} and for single top
($\tbbar$) production in the $W+\bbbar$
signature~\cite{single_top_CDF}.  Associated Higgs production via
$\ttbar H$ is expected to produce W+6 jets, of which 4 are b-quarks;
associated W and Z production via $\ttbar W$ or $\ttbar Z$ will also
produce W+6 jets, of which two will be b-quarks.

Precise measurements of the \WNjets~\cite{plus_only} and
 \ZNjets~\cite{Zgamma_poles} channels, where N is the number of jets,
 for values of N between 0 and at least 6, including the cases where
 pairs of the jets are either $\bbbar$ or $\ccbar$, would thus provide
 a broad search in a number of possible signatures of physics beyond
 the SM.  The importance of calculating the cross sections for these
 channels has long been
 recognized~\cite{theory0,theory1,theory2,theory3,theory4,theory5,theory6,
 theory7,theory8,theory9,theory10,theory11,theory12,theory13}; the
 development of sophisticated Monte Carlo programs capable of handling
 more particles in the final state at leading order (LO), or in some
 cases, next-to-leading order (NLO), now enables us to contemplate
 much more precise tests in the upcoming Tevatron Run II and at the
 LHC.

However, direct measurements of the production cross sections of
\WNjets or \ZNjets signatures suffer from inherent theoretical and
experimental uncertainties associated with the definition and
measurement (and hence counting) of jets. Among the dominant
experimental uncertainties are the energy response of the detector to
a jet (`energy scale'), additional energy contributions from the
underlying event (that part of the $\pbarp$ collision not directly
involved in the hard parton-parton collision that produces the W or
Z), backgrounds from misidentified non-electroweak events, and jet
acceptance.  These effects and others can change the number of jets
measured in a given event. Uncertainties in the theoretical SM
predictions are dominated by the choice of \Qsq~scale, the parton
distribution function (PDF), initial/final state radiation (ISR/FSR), and the
non-perturbative evolution of partons into on-shell particles that
would then be detected. All of these effects combined mean that the
measurement of a specific exclusive N-jet channel such as W+ 4 jets
will be completely dominated by systematic uncertainties at the
Tevatron in Run II and at the LHC.

In this note we use the Monte Carlo programs
MadGraph~\cite{1MadGraph,2MadGraph} and MCFM~\cite{MCFM1,MCFM2,MCFM3}
to explore using the measured ratios of \WNjets to \ZNjets
at each value of N to provide a much more precise test of the SM than
can be made by measuring the cross sections themselves~\cite{Maria}.
The W bosons are assumed to be identified by the leptonic decay $W^+
\rightarrow e+\nu$, and the $\Zg$ intermediate state by $\Zg
\rightarrow e^+e^-$. 
In most of the above models of new physics the production of new
particles decaying into W and \Zg + jet final states would change the
ratio from its SM prediction.  The uncertainties listed above, except
the misidentification backgrounds, are expected to cancel to a large
degree, and the backgrounds can themselves be made to partially cancel
by deriving the $\ellnu$ and $\ellpellm$ event samples from a common
inclusive high-$\Pt$ lepton sample~\cite{EtPt}.  We use data from the
CDF~\cite{CDF_Wjets} collaboration from Run I at the Fermilab Tevatron
to estimate the sensitivity to contributions from non-SM processes
using the \WNtoZN ratio method.

\section{\label{section:exp_uncertainty}Experimental Uncertainties in \WNtoZN}

The CDF Collaboration has published comprehensive studies of
inclusive~\cite{inclusive} \mbox{$W \rightarrow e \nu +N$ jets} and
\mbox{$\Zg \rightarrow e^+ e^- + N$ jets} production in $\pbarp$
collisions at $\roots = 1.8$ TeV~\cite{CDF_Wjets}.  The \Dzero
collaboration has measured the ratio of cross sections (W+ 1 jet)/(W+0
jet)~\cite{D0_Wjets}; as the \Dzero measurements are less extensive in
the number of jets (N) and do not include measurements of \Z+ jets, we
focus here on the CDF measurements.

The CDF W selection required an electron with $\Et > 20$ GeV and $|\eta|
< 1.0$, and missing transverse energy~\cite{EtPt} $\mett > 25$
\GeV. The Z selection required one electron satisfying the same
charged lepton requirements, and a second electron with $\Et>20$ \GeV
for $|\eta|<1.0$, $\Et>15$ \GeV for $1.1<|\eta|<2.4$, and $\Et>10$
\GeV for $2.4<|\eta|<3.6$. Jet identification~\cite{jet_id} was made
with a cone size in $\eta$-$\phi$ space of $\Delta R = 0.4$, a
threshold of $\Et \gt 15$ \GeV, and an $\eta$ range of $|\eta| \lt
2.5$. Multiple jets were required to be separated from each other in
$\eta$-$\phi$ space by a distance $\Delta R> 0.52$; the requirement
that the electron be `isolated' from other clusters of energy in the
calorimeter also corresponds to requiring $\Delta R> 0.52$ between the
electron and each jet~\cite{CDF_Wjets}.

\begin{figure}[!th]
\includegraphics[width=0.45\textwidth]{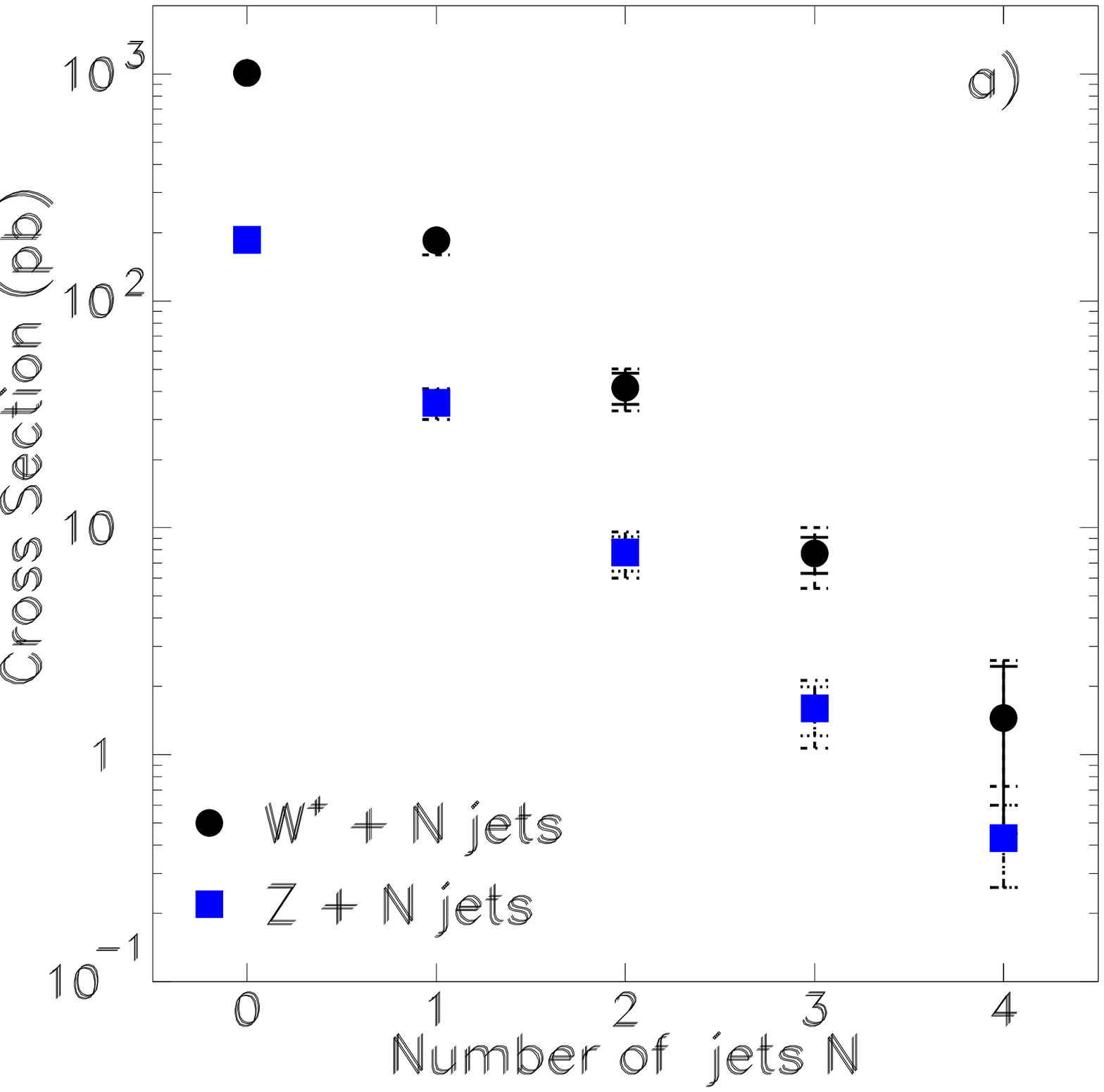}
\hfil
\includegraphics[width=0.45\textwidth]{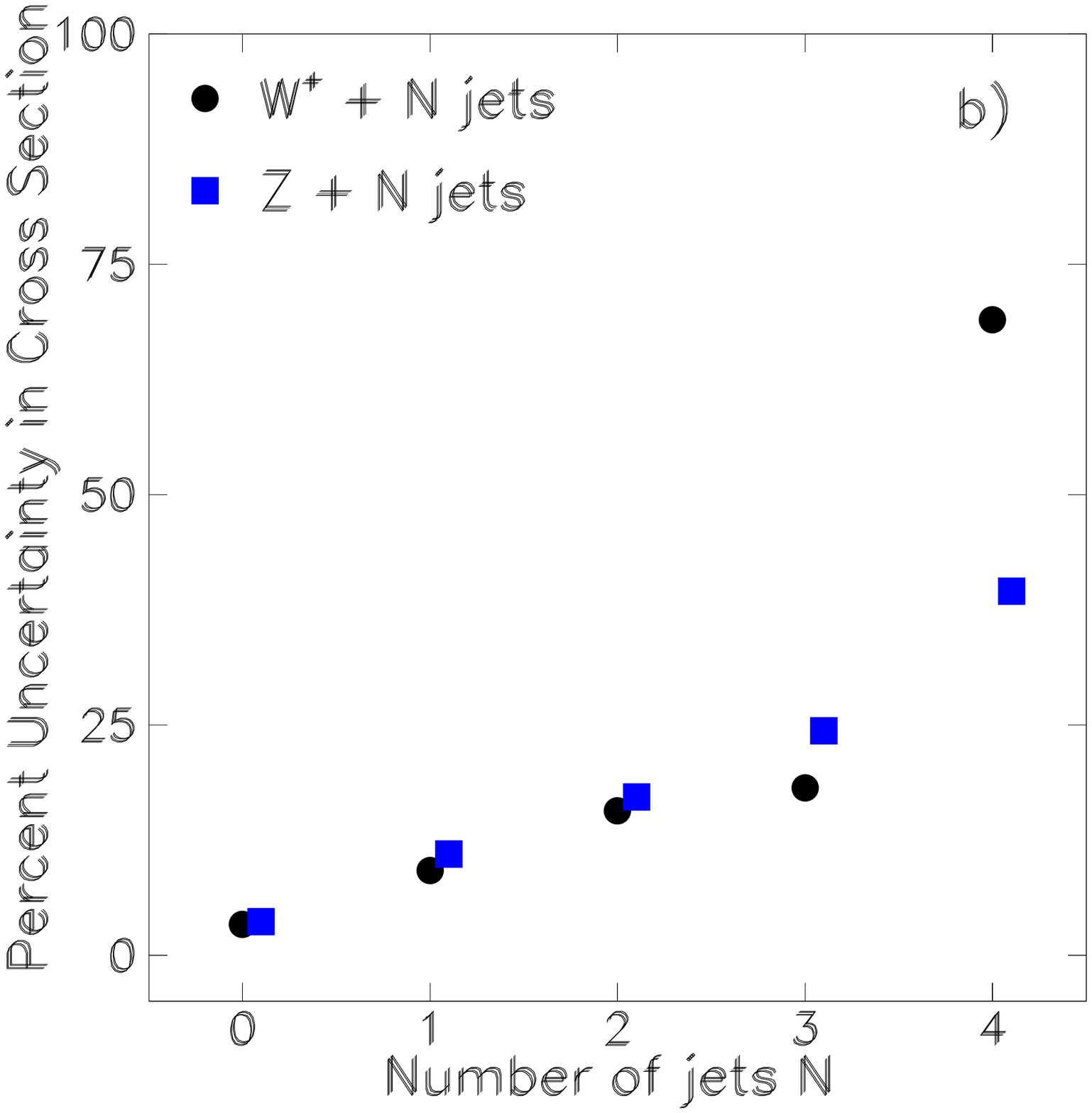}
\caption{ a) The measured cross sections for the signatures \WpenuNjets
and \ZeeNjets versus the number of jets, N, in $W^+$ and \Zg
production in $\pbarp$ collisions at $\roots=1.8$ \TeV. The data are
from the CDF~\cite{CDF_Wjets} collaboration and were originally
reported as inclusive cross sections.  In computing exclusive cross
sections from these, the uncertainties have been calculated in two
ways.  The dotted error bars were calculated assuming no correlations
(giving an upper bound for the uncertainty) and the solid error bars
were calculated assuming complete correlation (giving a lower
bound);
%The horizontal line
%is (half) the world average ~\cite{RPRD,RPRL,RDzero} of
%$R\equiv\sigma(W^+ + W^-)/\sigma(Z)$.
b) the percent uncertainty in the \WpNjets and \ZNjets
cross sections.  The uncertainties shown are the lower bounds
(corresponding to the solid error bars in plot a).  The figure
shows the rapid growth of the uncertainties with N, the number of jets.
}
\label{data_wandz}
\end{figure}

The individual (exclusive) cross sections extracted from the inclusive
cross sections measured by CDF for \WNjets and \ZNjets versus the
number of jets, N, are displayed in Table~\ref{CDFnumbers} and
Figure~\ref{data_wandz}, after being modified for comparison with
MadGraph's \Wp predictions by dividing the CDF cross sections for $W^+
+ W^-$ by two. The uncertainties have been calculated in two ways:
assuming no correlations (giving an upper bound for the uncertainty)
and assuming complete correlation (giving a lower bound).  The
uncorrelated uncertainties at each value of N have been calculated by
subtracting the uncertainties of higher values of N in quadrature, and
are reported first in the table. This overestimates the uncertainties,
but as the (N+1)th channel is typically only 20\% of the Nth channel
the overestimate is not large. The correlated uncertainties at each
value of N have been calculated by subtracting the uncertainties of
higher values of N; these uncertainties are reported second in the
table.
%The horizontal line is (half) the world average
%~\cite{RPRD,RPRL,RDzero} of $R\equiv\sigma(W^+ +W^-)/\sigma(Z)$.
% with dcolumn

%
% Table of CDF measured ratios
%
%\begingroup
%\squeezetable
%\begin{table}[!t]
\begin{table*}
\caption{\label{CDFnumbers}The cross sections times branching ratios
for \WpNjets and \ZNjets production (first two columns) extracted from
the CDF measurements versus the number of jets, at $\roots=1.8$
\TeV. These are used to calculate the ratios of the \WpNjets to \ZNjets
jet cross section times branching ratio (third column). Also shown are
the (less robust) ratios of \WNtoWNp and \ZNtoZNp.  The first
uncertainty given is the uncorrelated uncertainty, while the second
(in parentheses) is the
correlated uncertainty.  These uncertainties are derived as discussed in the
text.}
\begin{tabular}{|c|D{.}{.}{4.2}@{$\pm$}D{.}{.}{3.2}@{(}l|D{.}{.}{3.2}@{~$\pm$~}D{.}{.}{2.2}@{(}l|D{.}{.}{1.2}@{~$\pm$~}D{.}{.}{1.2}@{(}l|D{.}{.}{1.2}@{~$\pm$~}D{.}{.}{1.2}@{(}l|D{.}{.}{1.2}@{~$\pm$~}D{.}{.}{1.2}@{(}l|}
\hline
$N$ & \multicolumn{3}{|c|}{$\sigma_{W^{+}+Nj}$} & \multicolumn{3}{|c|}{$\sigma_{Z + Nj}$} & 
\multicolumn{3}{|c|}{$\sigma_{W^{+}+Nj}$/$\sigma_{Z + Nj}$}  & 
\multicolumn{3}{|c|}{$\sigma_{W+Nj}$/$\sigma_{W+N+1j}$}  &
\multicolumn{3}{|c|}{$\sigma_{Z+Nj}$/$\sigma_{Z+N+1j}$}  \\
\hline
0 & 1010 & 54 & 34) & 185.8 & 11.1 & 6.7) & 5.43 & 0.44 & 0.27) & 5.46
& 0.78 & 0.53) & 5.23 & 0.87 & 0.60)  \\
%  & & 34 & & 6.7 & & 0.27 & & 0.53 & & 0.60 \\

1 & 185 & 25 & 17) & 35.5 & 5.5 & 3.9) & 5.21 & 1.06 & 0.75) & 4.46 &
1.10 & 0.81) & 4.55 & 1.26 & 0.93)  \\
%  & & 17 & & 3.9 & & 0.75 & & 0.81 & & 0.93  \\

2 & 41.5 & 8.7 & 6.5) & 7.8 & 1.8 & 1.34) & 5.32 & 1.65 & 1.23) & 5.42
& 1.98 & 1.30) & 4.88 & 1.97 & 1.46)  \\
%  & & 6.5 & & 1.34 & & 1.23 & & 1.30 & & 1.46  \\
 
3 & 7.7 & 2.3 & 1.4) & 1.6 & 0.53 & 0.39) & 4.78 & 2.14 & 1.46) & 5.28
& 4.48 & 3.77) & 3.72 & 1.92 & 1.73)  \\
%  & & 1.4 & & 0.39 & & 1.46 & & 3.77 & & 1.73  \\
 
4 & 1.45 & 1.15 & 1.00) & 0.43 & 0.17 & 0.17) & 3.37 & 2.99 & 2.68) & \multicolumn{3}{|c|}{-} & \multicolumn{3}{|c|}{-}  \\
%  & & 1.00 & & 0.17 & & 2.68 & \multicolumn{2}{|c|}{-} & \multicolumn{2}{|c|}{-}  \\
 
%5 & \multicolumn{2}{|c|}{} & \multicolumn{2}{|c|}{} &
%\multicolumn{2}{|c|}{} & \multicolumn{2}{|c|}{} & \multicolumn{2}{|c|}{}  \\ 
%  & \multicolumn{2}{|c|}{} & \multicolumn{2}{|c|}{} &
%\multicolumn{2}{|c|}{} & \multicolumn{2}{|c|}{} & \multicolumn{2}{|c|}{} \\

\hline
\end{tabular}
\end{table*}
%\endgroup

The estimated CDF systematic uncertainties are broken down according
to the source of each uncertainty in Table~\ref{systematics} versus
the inclusive number of jets.  One can see that in general the quoted
systematic uncertainties grow rapidly with N, as described in detail
in Ref.~\cite{CDF_Wjets} This is due to the difficulties of counting
jets given the rapidly falling spectrum in $\Et$ and the uncertainties
in measuring the energy of a jet, and, to a lesser extent,
uncertainties in the position of the jet with respect to the limit in
$\eta$ in the jet selection.  In addition, energy deposited in the
calorimeter from the fragments of the collision not directly produced
by the `hard' interaction that produced the boson, called the
`underlying event', contribute to the total energy measured in the jet
cone, and can promote jets from below threshold to over threshold,
changing the number of jets in the event. Similarly, multiple
interactions from separate $\pbarp$ collisions in the same bunch
crossing~\cite{multiple_interactions} can contribute energy in the jet
cone. There are smaller contributions from uncertainties in the
acceptance for the leptons, for the `obliteration' of a lepton by a
jet (if a jet lands close to a lepton the lepton can fail the
identification criteria), and uncertainties in the contribution from
decays of the top quark. Lastly, the uncertainty due to backgrounds
from processes other than vector boson production (`QCD background')
grows with the number of jets.

\begin{table*}
%\begin{table}[!htb]
\caption{\label{systematics}The systematic uncertainties in percent on
the measured CDF inclusive $W+N$ jet production cross sections for for
N=1 to N=4 (column 1)~\protect{\cite{CDF_Wjets}}. The successive
columns are the uncertainties in the cross sections due to
uncertainties in: the calorimeter jet energy scale, the underlying
event, QCD background to W identification, multiple $\pbarp$
interactions in a single event, the value of the maximum allowed
$|\eta|$ for jets to be counted, the W acceptance, `obliteration' of
an electron by the superposition of a jet, and contributions from the
top quark.  The larger error bar is quoted in the case of asymmetric
uncertainties. }
\begin{tabular}{|c|r|r|r|r|r|r|r|r|}
\hline
N(Jets) & $Et_J$ Scale & Und Ev & QCD Bkgd & Mult Int & $\eta_J$ & Acc & Oblit & Top \\
\hline
$\ge 1$ & 6.8\% & 5.8\% & 5.2\%  & 3.2\% & 1.9\% & 0.8\% & 0.2\% & 0.05\% \\
$\ge 2$ &  11\% & 9.8\% & 5.4\%  & 7.2\% & 3.7\% & 1.0\% & 0.3\% & 0.3\% \\
$\ge 3$ &  17\% & 16\%  & 9.1\%  & 9.8\% & 4.8\% & 1.8\% & 0.6\% & 1.3\% \\
$\ge 4$ &  23\% & 21\%  & 15.8\% & 14\%  & 5.5\% & 3.5\% & 1.3\% & 0.5\% \\
\hline
\end{tabular}
\end{table*}

%
% CDF data
%

\begin{figure*}[!th]
\includegraphics[width=0.45\textwidth]{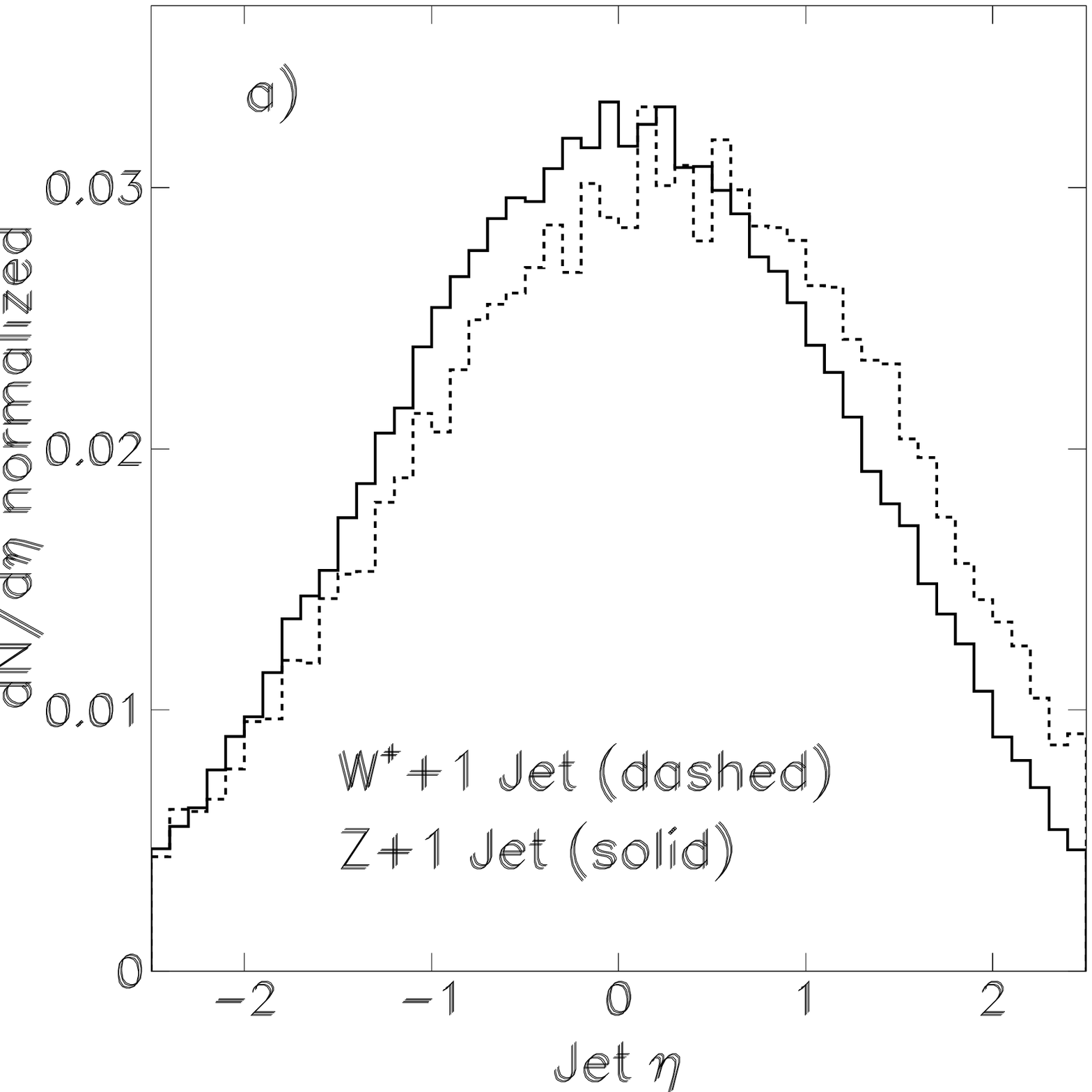}
\hfil
\includegraphics[width=0.45\textwidth]{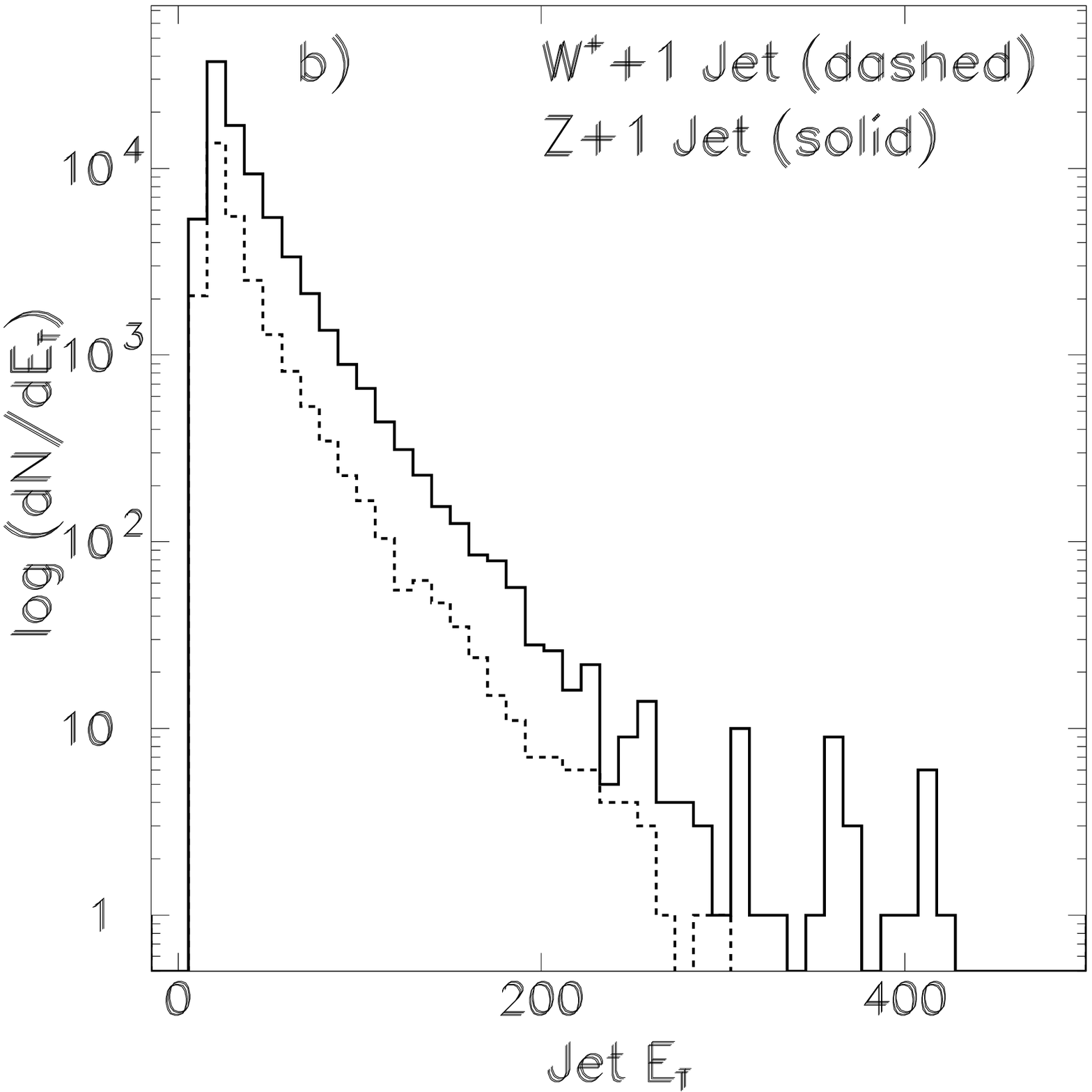}
\caption{\label{Madgraph_jet_et_eta}The plot on the left (a) shows the
(normalized) jet $\eta$ distributions for \Wpenujet (dashed) and
\Zeejet (solid) events satisfying the selection criteria described in
the text.  The plot on the right (b) shows the corresponding jet $\Et$
distributions, log ($dN/d\Et$) versus N.  Both plots are predictions
at LO using MadGraph~\protect{\cite{2MadGraph}}. Uncertainties in the
ratios $\WpNtoZN$ due to the uncertainty in the jet rapidity cut at
$\eta$=2.5 are estimated from the shapes in the left-hand plot, and
those due to the uncertainty in the jet energy scale from the
right-hand plot.}
%The ratio of the two is shown as a solid line; the ratio after initial
%and final state radiation as generated by Pythia~\cite{Pythia} is
%shown as a dotted line.
\end{figure*}

%The quoted uncertainties on the cross sections measured by
%CDF
\enlargethispage{0.5in} The largest uncertainty is from the jet energy
scale. This uncertainty will cancel in the production of W+jets and
\Zg + jets events to the extent that the spectra in $\Et$, the
distribution in $\eta$, and the composition (e.g. quark versus gluon)
of the jets in the two processes are the same~\cite{apologia}.
Figure~\ref{Madgraph_jet_et_eta} shows the spectra in $\eta$ and $\Et$
generated with the MadGraph Monte Carlo program ~\cite{2MadGraph} at
LO. Using the difference of the ratio of the fitted slopes of the
$\Et$ distributions for W and Z production in
Figure~\ref{Madgraph_jet_et_eta} times a typical uncertainty in the
$\Et$ scale of 20\%~\cite{CDF_Wjets} at 20 \GeV gives an estimate of
the uncertainty in the ratio of 2\%. The effect of the finite
acceptance in $\eta$ for jets depends on the difference in the
distributions in $\eta$ of jets in W or Z production; taking the
difference shown in Figure~\ref{Madgraph_jet_et_eta} times the
estimated variation with rapidity in jet response~\cite{jet_balancing}
gives an estimate of the uncertainty in the ratio of 1\%.
% as well as the effects of initial state
%radiation as modelled by Pythia~\cite{Pythia}. The ratios of jets in W
%events to those in \Zg events are shown as solid lines; the ratios
%after initial and final state radiation as generated by
%Pythia~\cite{Pythia} are shown as dotted lines.  

%The spectra are very
%similar, and so within the limited scope of this note we consider the
%effects of the jet energy scale to cancel~\cite{jet_balancing}. The
%effect of the $\eta$ range selection on the jets we expect to cancel
%similarly.

%
% Figure 3- the jet eta and et distributions from MadGraph
%

%\clearpage

The second largest systematic uncertainty is from the effects of
energy from the underlying event, which can `promote' a 3-jet event to
being a 4-jet event, for example, by boosting a lower-energy jet above
the jet-counting threshold in $\Et$. We expect that the underlying
events in W and in Z events should be very similar; studies of the
underlying event in jet events~\cite{RField} predict that the
contribution from the beam fragments, which could be different due to
the different quark diagrams in W and in Z production, are a small
portion of the total. However, the energy per tower contributed by the
underlying event, and hence the effect on `promotion' of jets, can be
directly measured in \WNjets and \ZNjets events. We consequently
assume that this uncertainty will be negligible in the ratio.

For higher ($\ge$ 4) jet multiplicities QCD backgrounds become
comparable to each of the above. The backgrounds in the $\Zg$ channel
are at the few percent level, and are measurable (and hence subtractable)
by counting same-sign events. Previous studies of the backgrounds to
inclusive W production by CDF~\cite{Sacha_thesis} for selection
criteria similar to those used here have shown that the background is
dominated by approximately equal contributions from leptons from heavy
flavor production and misidentified hadrons. How well these can be
measured with the new Run II detectors is not yet known; the former can be 
measured with the silicon vertex detectors, and the latter can be
measured by conventional background techniques.

The next largest systematic uncertainty in the Run I CDF cross section, 
contributions from multiple $\pbarp$ interactions, should cancel
identically in the ratio, as it is uncorrelated with the hard
scattering.

The remaining uncertainties due to acceptance, `obliteration' of a
lepton by a jet, and contributions from top decay, are at most at the few
percent level~\cite{CDF_Wjets}.

\section{\label{section:monte_carlo}Monte Carlo Programs and Event Selection Criteria}
\label{selection}

We have explored the W to \Z ratios in $\pbarp$ collisions at the
Tevatron energy of $\roots = 1.96$ TeV using the Monte Carlo programs
MadGraph~\cite{2MadGraph} and MCFM~\cite{MCFM3}.  Samples of
\WpenuNjets and \ZeeNjets, for N up to 4, were produced at LO using
MadGraph. MCFM was used to explore the ratios for up to 2 jets at NLO,
and to understand the dependence of the ratios on the \Qsq~scale and
on the parton distribution functions for up to 4 jets at LO. Jets are
treated at the `parton level' with kinematic selections applied to the
4-vectors with no fragmentation or detector simulation.

We consider only the production in first-order electroweak processes
of the W+jets and Z+jets channels -- i.e. production of boson + jets
from the WW, WZ, and ZZ channels are excluded. We also exclude
$\ttbar$ and $\tbbar$ production; the method proposed here should
allow a more precise determination of the non-top W+jets production,
the dominant background in the top channel, and hence should allow
more precise measurements of the top quark mass and cross section.

The selection criteria and strategy for W and \Zg events used in the
Monte Carlo studies were developed for the measurement of $R$, the
ratio of inclusive cross sections $R \equiv\sigma
(W)/\sigma(\Zg)$~\cite{R}. To minimize systematic uncertainties in the
ratio due to the trigger and lepton selection, both W and \Zg events
are selected from a common sample of inclusive central high transverse
momentum~\cite{EtPt} ($\Pt$) leptons, with transverse energy ($\Et$)
greater than 25 \GeV and pseudo-rapidity $(|\eta|$) less than 1.0.
The second lepton from the boson decay, either another charged lepton
(from \Zg decay) or a neutrino (from W decay), is required to have
$\Et> 25$ GeV; in the neutrino case this is implemented by requiring
the missing transverse energy ($\mett$) to be greater than 25 GeV.

Jets are required to have $\Et> 15$ \GeV and to be within $|\eta| <
2.5$. Our MC studies are at parton level, so that there are no
considerations of cone size, energy scale, or acceptance corrections
in the Monte Carlo numbers.  

% \begin{table}%[H] add [H] placement to break table across pages
% \begin{ruledtabular}
% Lines of table here ending with \\
% \end{tabular}
% \end{ruledtabular}
% \end{table}

%
% Table of MC predicted ratios
%
%\begingroup
%\squeezetable
\begin{table*}

\caption{\label{predicted_numbers}MadGraph leading order predictions
of cross sections times branching ratio for \WpNjets and \ZNjets
production (first two columns) versus the number of jets, at
$\roots=1.96$ \TeV, which are used to calculate the ratios of the
\WpNjets to \ZNjets jet cross section times branching ratio (third
column). Also shown are the (less robust) ratios of \WNtoWNp and
\ZNtoZNp (last two columns).}

%\begin{ruledtabular}
\begin{tabular}{|c|D{.}{.}{3.3}@{~$\pm$~}D{.}{.}{1.3}|D{.}{.}{2.3}@{~$\pm$~}D{.}{.}{1.3}|D{.}{.}{1.2}@{~$\pm$~}D{.}{.}{1.2}|D{.}{.}{1.2}@{~$\pm$~}D{.}{.}{1.2}|D{.}{.}{1.2}@{~$\pm$~}D{.}{.}{1.2}|}
\hline
$N$ & \multicolumn{2}{|c|}{$\sigma_{W^{+}+Nj}$} &
\multicolumn{2}{|c|}{$\sigma_{Z + Nj}$} &
\multicolumn{2}{|c|}{$\sigma_{W^{+}+Nj}$/$\sigma_{Z + Nj}$}  &
\multicolumn{2}{|c|}{$\sigma_{W+Nj}$/$\sigma_{W+N+1j}$}  &
\multicolumn{2}{|c|}{$\sigma_{Z+Nj}$/$\sigma_{Z+N+1j}$}  \\
\hline

0 & 341.5 & 0.5 & 67.0 & 0.2 & 5.10 & 0.02 & 8.11 & 0.06 & 6.41 & 0.02  \\

1 & 42.1 & 0.3 & 10.45 & 0.01 & 4.03 & 0.03 & 5.08 & 0.05 & 4.91 & 0.07  \\

2 & 8.28 & 0.05 & 2.13 & 0.03 & 3.89 & 0.06 & 4.93 & 0.07 & 4.75 & 0.09  \\ 

3 & 1.68 & 0.02 & 0.448 & 0.006 & 3.75 & 0.07 & 4.71 & 0.09 & 5.15 & 0.09  \\ 

4 & 0.357 & 0.005 & 0.087 & 0.001 & 4.10 & 0.07 & %8.71 & 0.24?
 \multicolumn{2}{|c|}{-} &
\multicolumn{2}{|c|}{-}  \\ 

%5 & 0.041 & 0.001? & \multicolumn{2}{|c|}{} & \multicolumn{2}{|c|}{} &
%\multicolumn{2}{|c|}{} & \multicolumn{2}{|c|}{}  \\ 

\hline
\end{tabular}
%\end{ruledtabular}
\end{table*}
%\endgroup

\section{\label{section:predicted_ratios_WZ}The Predicted Ratios \WpNtoZN}

\begin{figure}[!ht]
\includegraphics[width=0.45\textwidth]{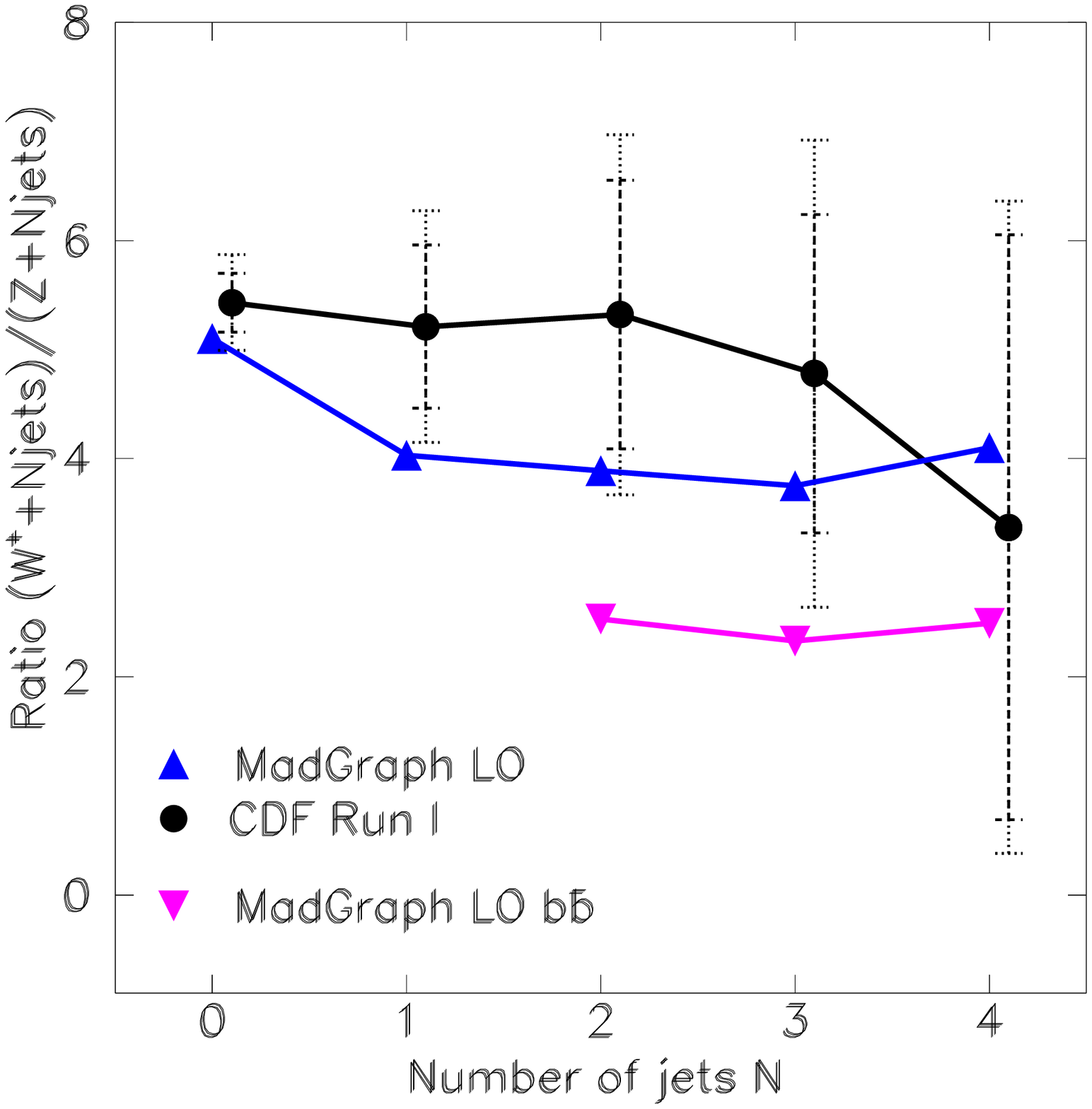}
\caption{\label{WNjet_to_ZNJet}A comparison of CDF data (circles)
with MadGraph (LO) predictions (triangles) for the ratio of
production cross sections times leptonic branching ratios for the
signature \WpNjets to the signature \ZNjets, \WpNtoZN, versus the
number of jets, N, in $W^{+}$ and \Zg production at $\roots=1.8$ \TeV
for the data and at $\roots=1.96$ \TeV for the predictions. The
case where two of the jets are b-quark jets are also shown (inverted
triangles). The statistical uncertainties on the predictions are
smaller than the symbols.}

\end{figure}

The predicted LO ratios \WpNtoZN are presented in
Figure~\ref{WNjet_to_ZNJet} and in Table~\ref{predicted_numbers}. To
determine the final uncertainty on this ratio will take a full
analysis of the Run II data set; in lieu of this we have made some
simple assumptions to get an estimate of the sensitivity in
cross-section for non-SM physics in each of the N-jet channels in Run
II of the Tevatron. We assume that the jet energy response of the
calorimeter will largely cancel for jets in \Zg events and W events as
discussed below. We also assume the effects of the underlying event in
\Zg and W events will similarly cancel. These are the two largest
contributors to the systematic uncertainties quoted in
Ref.~\cite{CDF_Wjets}.

%
% Figure 1- W to Z ratios
%

\section{\label{section:predicted_ratios_nton+1}The Predicted Ratios 
\WNtoWNp and \ZNtoZNp}

\begin{figure*}
\includegraphics[angle=0,width=0.45\textwidth]{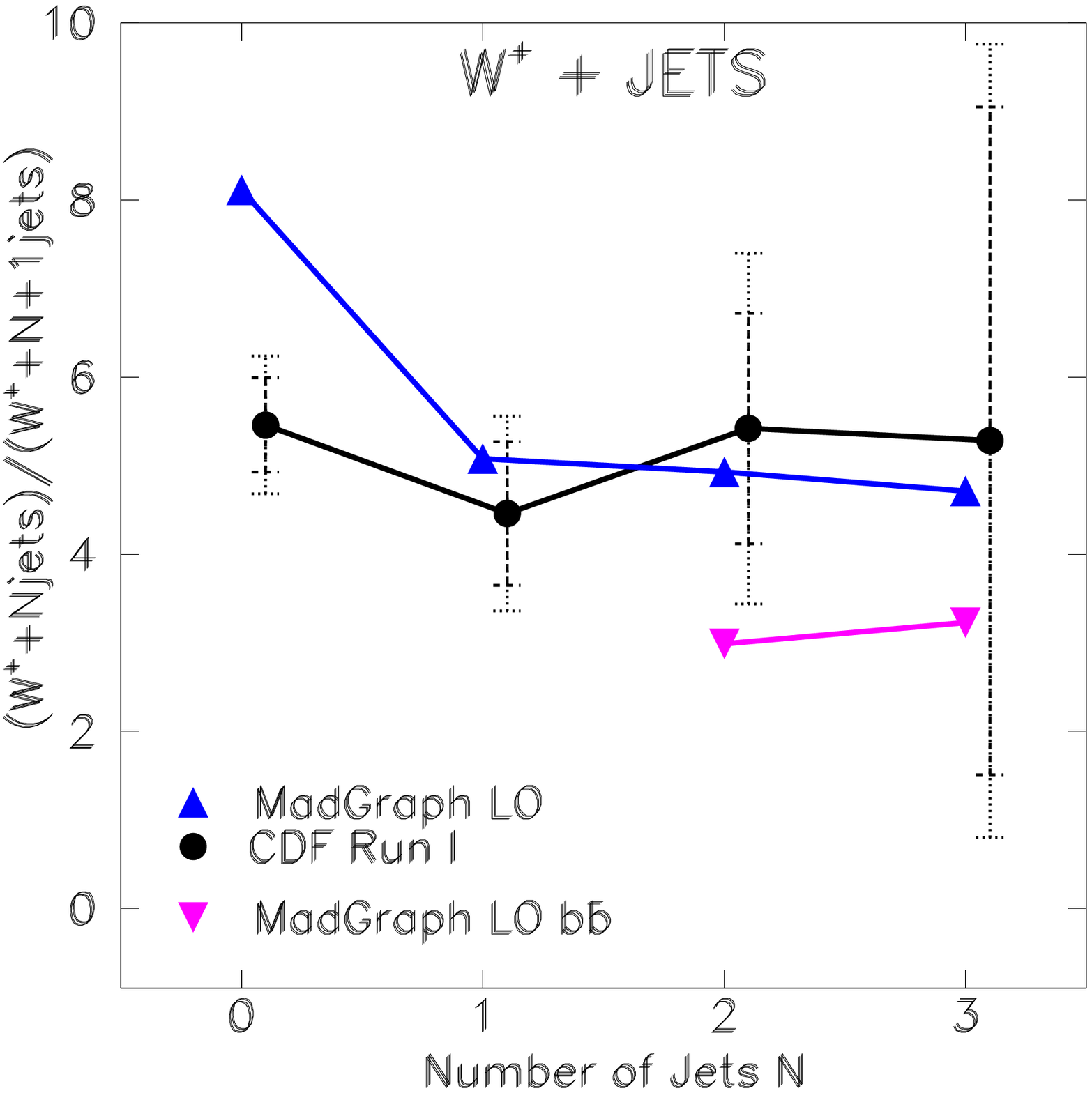}
\hfil
\includegraphics[angle=0,width=0.45\textwidth]{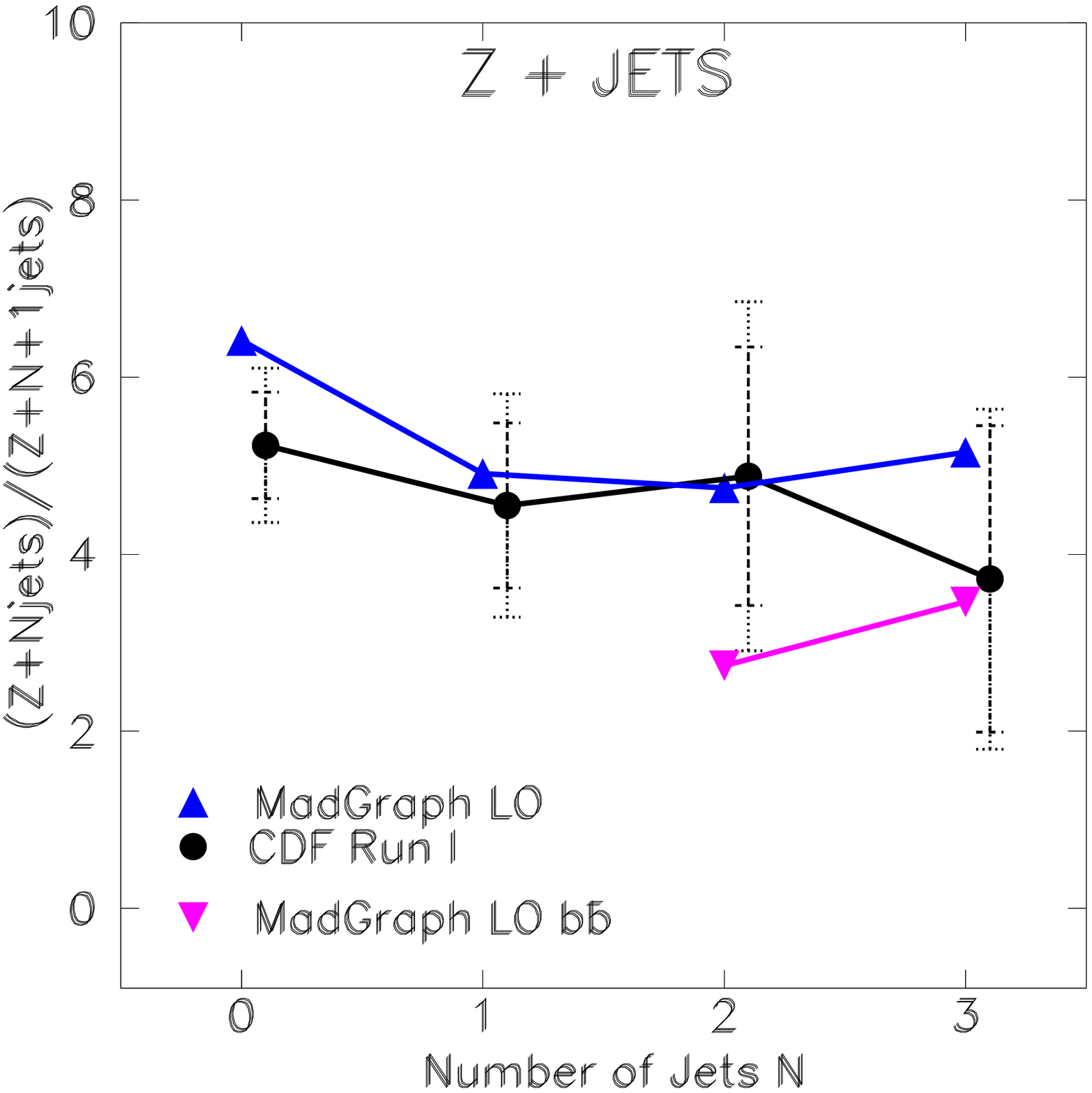}
\caption{\label{Njet_to_N+1Jet}The ratio of cross sections times
branching ratios, $\WpNtoWpNp$ (left hand plot) and $\ZNtoZNp$ (right hand
plot) versus the number of jets, N, in W and Z production at
$\roots=1.8$ \TeV for the data and at $\roots=1.96$ \TeV for the
predictions. The data (circles) are from the CDF~\cite{CDF_Wjets} and
\Dzero~\cite{D0_Wjets} collaborations; the predictions are at leading
order from MadGraph~\cite{2MadGraph}.
% and next-to-leading order from MCFM~\cite{MCFM}. 
The MadGraph cross sections for when the jets are from gluons
or light quarks are shown with triangles, while inverted triangles represent
when two of the jets are from b-quarks.}
\end{figure*}

While the ratios of cross sections \WNtoWNp and \ZNtoZNp are much more
difficult to measure precisely than the \WpNtoZN ratios, we include
the generator-level LO predictions for them here as they are often
used in extrapolations in $N$ to estimate backgrounds at large N, and
also to measure the strong interaction coupling.  These are reported
in Table~\ref{predicted_numbers}, and shown in
Figure~\ref{Njet_to_N+1Jet}.

%
% Figure 2-  ratios within the W's and within the Z's 
%        

%\begin{figure}[!ht]

%\clearpage

\section{\label{section:theor_uncertainty}Theoretical  Uncertainties in \WpNtoZN}

The two largest uncertainties in the predicted LO \WNjets and \ZNjets
cross sections are expected to be due to choice of \Qsq~scale and parton
distribution function (PDF).  We investigate the dependence of the
ratio on these two choices using MCFM.
%In addition, we examine the
%difference in the leading order predictions versus those at next-to-leading
%order for a small number of jets using MCFM.

\subsection{\label{subsection:q}Dependence on the \Qsq~Scale}

\begin{figure}[!h]
\includegraphics[width=0.45\textwidth]{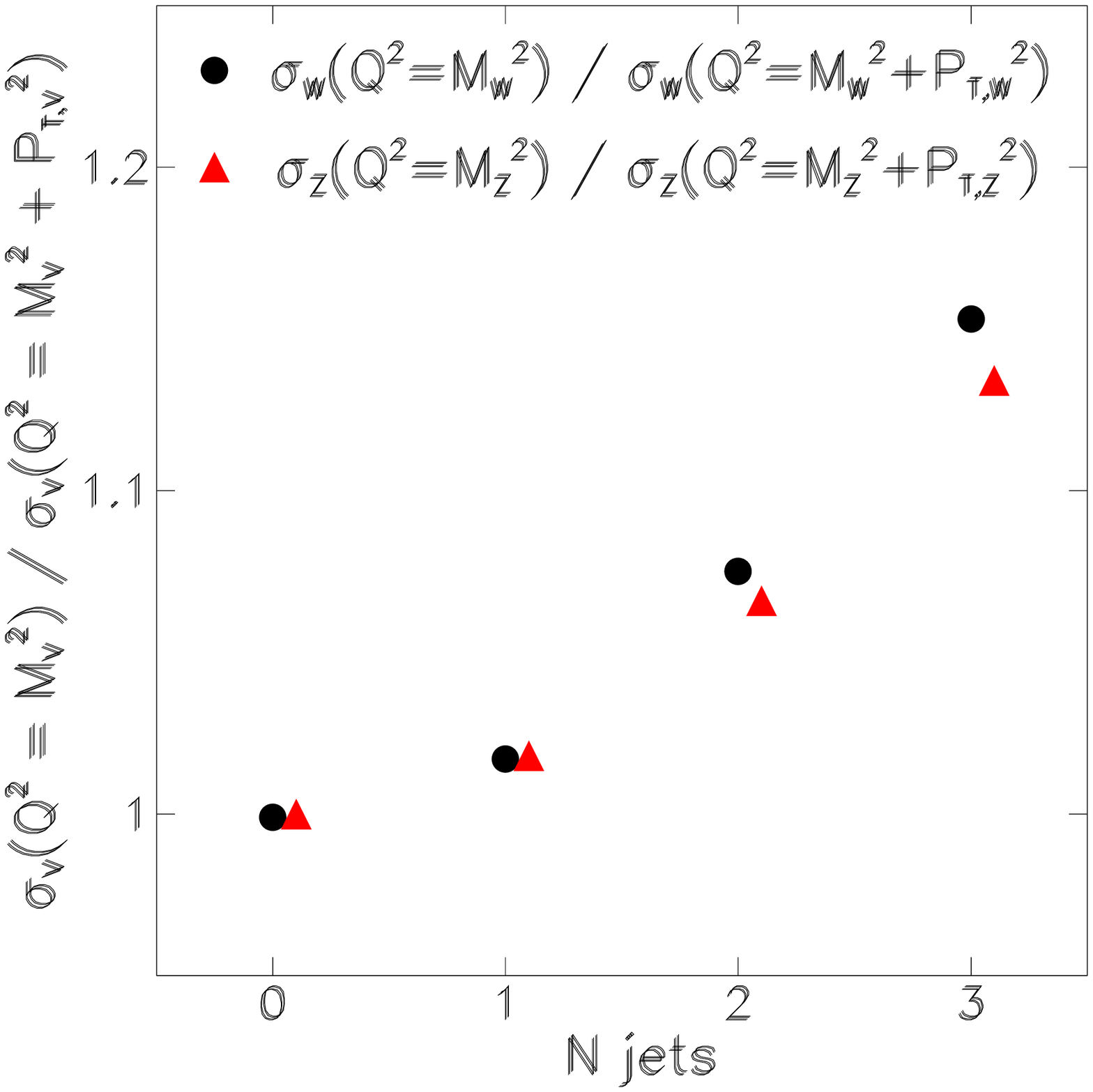}
\caption{\label{mcfm_vq1_to_vq2}The ratios $\sigma$(\WpNjets at \QsqeqM)
to $\sigma$(\WpNjets at \QsqeqMP) and $\sigma$(\ZNjets at \QsqeqM)
to $\sigma$(\ZNjets at \QsqeqMP).  Changing the \Qsq~scale
significantly changes the cross sections, by up to approximately 15\%.
However the ratio of W to Z cross sections changes much less (see
Table~\ref{qstudy}).}
\end{figure}

\begin{figure}[!h]
\includegraphics[width=0.45\textwidth]{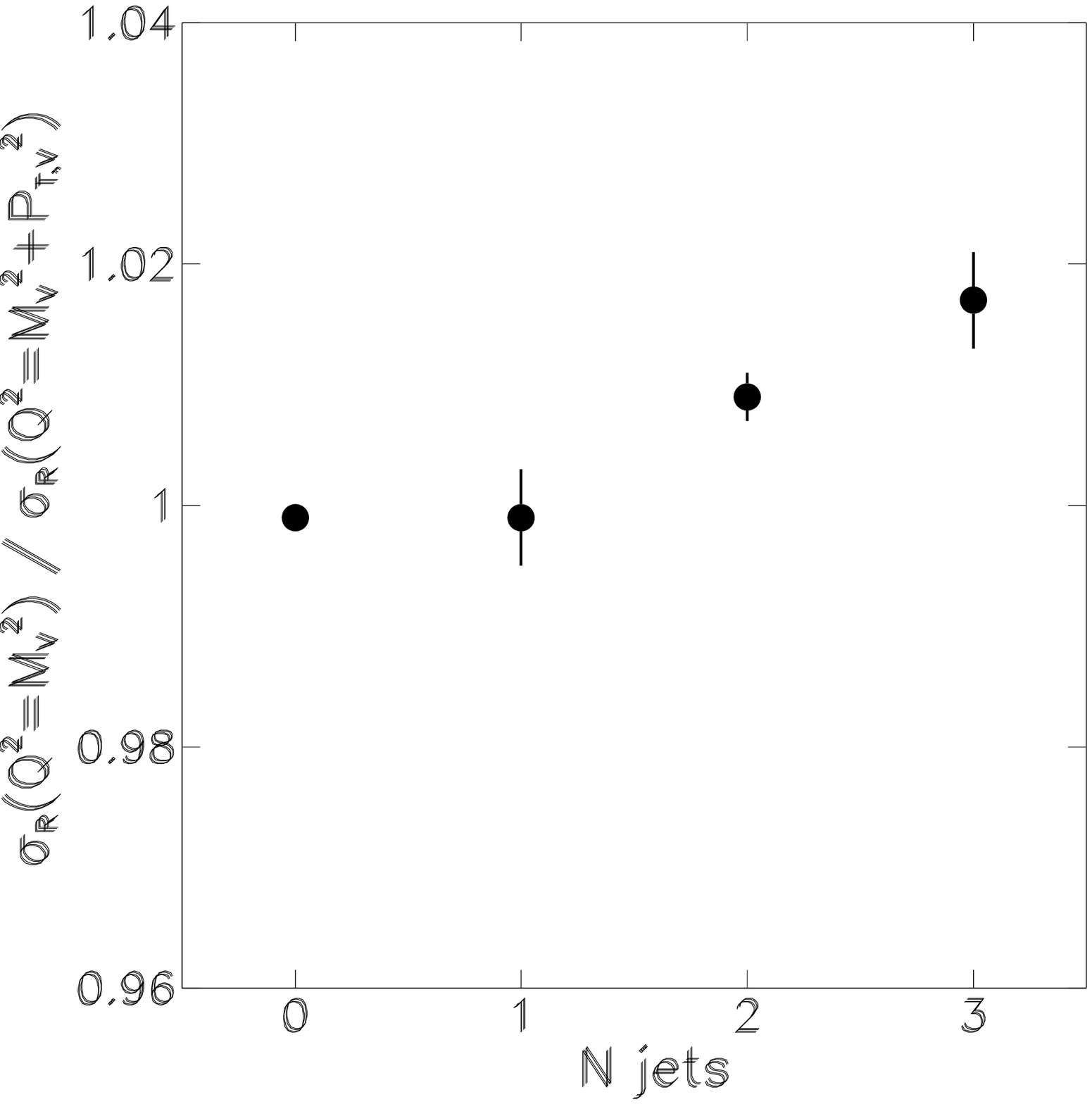}
\caption{The ratios ($R^+$ at \QsqeqM) to
($R^+$ at \QsqeqMP), where $R^+ ~=$   \WpNtoZN.  Changing the
\Qsq~scale affects this ratio by $\sim 2\% $ while the individual
cross sections change by more than $15\%$.}
\label{mcfm_qstudy_ratiowz}
\end{figure}

The effect of the choice of \Qsq~scale is expected to partially cancel
in \WNjets and \ZNjets production, as both proceed through a
Drell-Yan-like process.  We define $\sigmaWN \equiv \sigma\WpNjets$
evaluated at $Q^2$, and, similarly, $\sigmaZN \equiv \sigma\ZNjets$.
The ratios of W and Z cross sections evaluated at $\QsqeqM$ and at
$\QsqeqMP$, $\sigmaWNm/\sigmaWNpm$ and $\sigmaZNm/\sigmaZNpm$, are
given in Table~\ref{qstudy} and shown in
Figure~\ref{mcfm_vq1_to_vq2}. Changing the \Qsq~scale affects the W
cross sections by as much as 15\% and affects the Z cross sections by
as much as 12\%.  However, changing the \Qsq~scale has much less
effect on the predicted ratio \WpNtoZN, which changes less than 2\%,
as shown in Figure~\ref{mcfm_qstudy_ratiowz} and in Table ~\ref{qstudy},
where the W/Z ratios evaluated at the two different values of $Q^2$
also are listed.

% mcfm qstudy predictions
%\begingroup
%\squeezetable
\begin{table*}
%\begin{table}[!htbp]
%\caption{\label{qstudy_w_njets}A comparison of MCFM-predicted W+jet
%cross sections in pb (for ${\text W}^+$ only) for ${\text Q}^2 =
%{\text W_{mass}}^2$ versus ${\text Q}^2 = {\text W_{mass}}^2 + {\text
%P_{T,W}}^2$,\\ with other selection criteria as in the text.}
\caption{\label{qstudy}MCFM predictions for the ratios \WpNjets and
\ZNjets with different \Qsq~scales (columns one and two), and the
ratio $R^+$ with different \Qsq~scales.  $R^+$ is \WpNtoZN, and 
\Qonesq corresponds to
\QsqeqM , while \Qtwosq corresponds to \QsqeqMP.}
%\begin{tabular}{|c||D{.}{.}{1.3}@{~$\pm$~}D{.}{.}{0.3}|D{.}{.}{1.3}@{~$\pm$~}D{.}{.}{0.3}|D{.}{.}{1.3}@{~$\pm$~}D{.}{.}{0.3}|}
\begin{tabular}{|c||r@{~$\pm$~}l|r@{~$\pm$~}l|r@{~$\pm$~}l|}
%\hline\hline
%\multicolumn{9}{|c|}{\Large\bf  W+Jets Cross Sections (pb)} \\
\hline
$N_{jets}$ & \multicolumn{2}{|c|}{$\sigma\text W^{+}$(\Qonesq) / $\sigma\text W^{+}$(\Qtwosq)} & 
\multicolumn{2}{|c|}{$\sigma\text \Z$(\Qonesq) / $\sigma\text \Z$(\Qtwosq)} &
\multicolumn{2}{|c|}{$\text R^{+}$(\Qonesq) / $\text R^{+}$(\Qtwosq)}  \\
\hline\hline
0 & ~~~~~~~0.999 & .001 & ~~~~~~1.000 & .001 & ~~~~0.999 & .001 \\
1 & 1.017 & .003 & 1.018 & .002 & 0.999 & .002 \\
2 & 1.075 & .002 & 1.066 & .002 & 1.009 & .002 \\
3 & 1.153 & .004 & 1.134 & .002 & 1.017 & .004 \\
%0 & 341.5 & 0.5 & 343 & 2 & 1.004 & 0.006 \\
%1 & 42.1 & 0.3 & 45.4 & 0.3 & 1.078 & 0.010 \\
%2 & 8.28 & 0.05 & 9.26 & 0.09 & 1.118 & 0.013 \\
%3 & 1.68 & 0.02 & 1.99 & 0.03 & 1.185 & 0.023 \\
%4 & 0.357 & 0.005 & 0.428 & 0.007 & 1.199 & 0.026 \\
\hline
%2 j  & 2 b & (4,2) & 0.0917 & 0.0008 & & & & \\
%3 j  & 2 b & (4,2) & 0.0307 & 0.0003 & & & & \\
%4 j  & 2 b & (4,2) & 0.0095 & 0.0003 & & & & \\
%4 j  & 2 b & (2,4) & 0.1983 & 0.0005 & & & & \\
%4 j  & 2 b & (4,4) & 0.210 & 0.002 & & & & \\
%\hline
%4 j  & 4 b & (4,2) & (4.1 & 0.1)E-5 & & & & \\
%\hline\hline
\end{tabular}
\end{table*}
%\endgroup
%\clearpage

%
% mcfm qstudy plot q1/q2 ratio
%

%
% mcfm qstudy plot w/z ratio
%

\subsection{\label{subsection:pdf}Dependence on the Choice of Parton Distribution Function}

\begin{figure}[!t]
\includegraphics[width=0.45\textwidth]{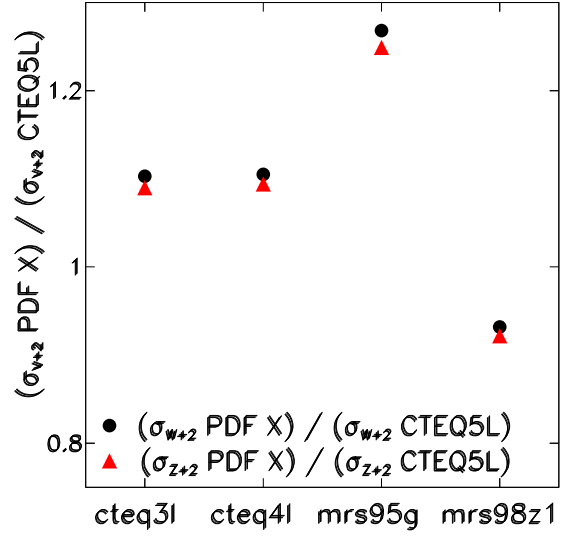}
\caption{\label{mcfm_pdfx_to_cteq5l}The ratios $\sigma$(W+2jets for
PDF X) to $\sigma$(W+2jets for CTEQ5L) and $\sigma$(Z+2jets for PDF X)
to $\sigma$(Z+2jets for CTEQ5L).  Changing the PDF affects the cross
sections quite significantly, by up to approximately 25\%. However the
ratio of W to Z cross sections changes much less (see Table~\ref{pdfstudy}).}
\end{figure}

\begin{figure}[htb]
\includegraphics[width=0.45\textwidth]{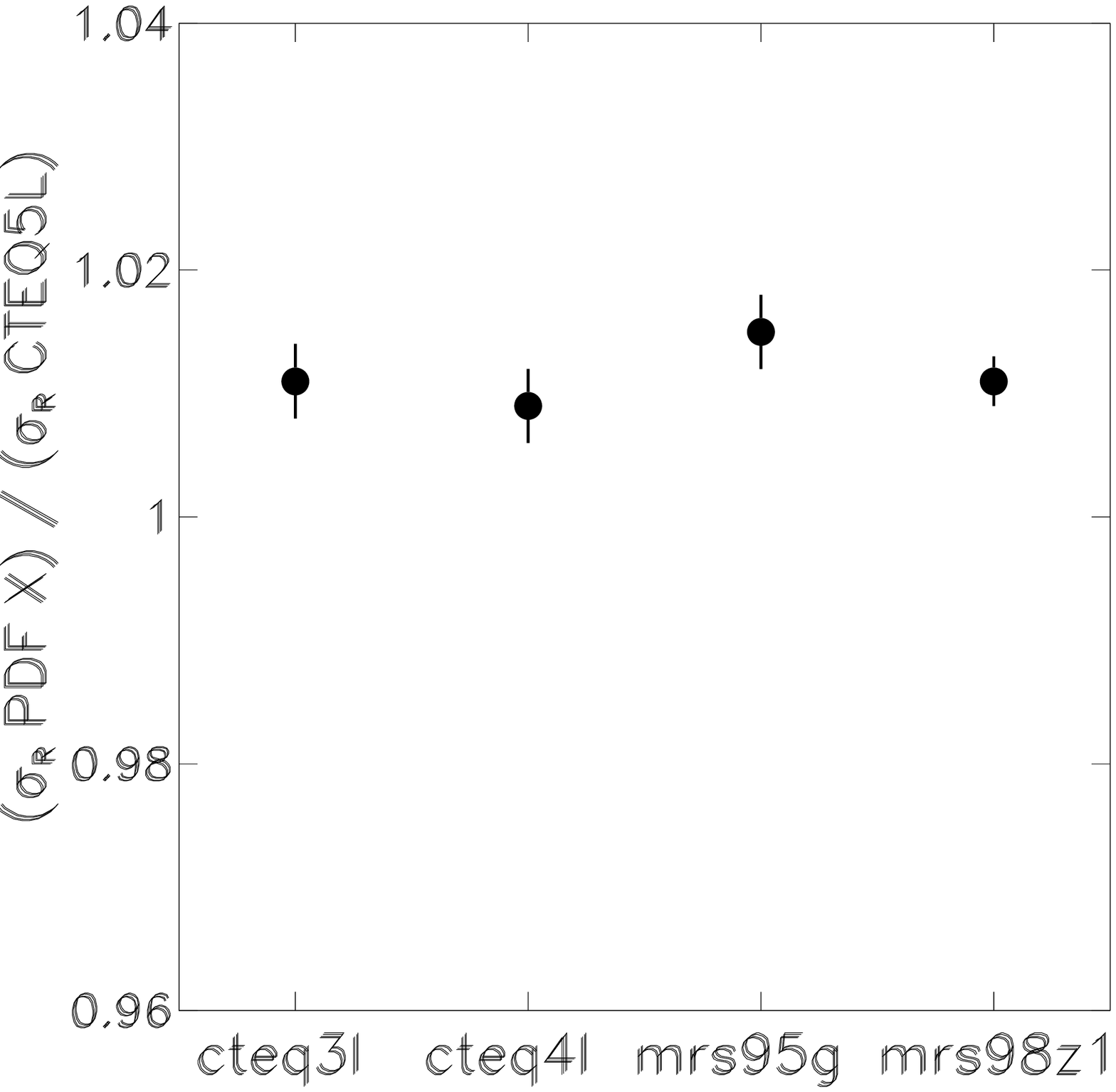}
\caption{\label{mcfm_pdfstudy_ratiowz}The ratios R for PDF X to R for
CTEQ5L, where R =~\WNtoZN.  Changing the PDF affects this ratio much
less - by at most 2\% - than it affects the individual cross sections.}
\end{figure}

 We have used the MCFM generator and a selection of parton distribution
 functions to investigate the dependence of the cross sections in the
 $W^{+}$+2jets and $\Z$+2jets channels.  The cross sections calculated with
 the CTEQ3L, CTEQ4L~\cite{CTEQ5}, MRST98~\cite{MRS98}, and
 MRSG95~\cite{MRS95} distributions were compared to the results
 calculated with CTEQ5L, the default PDF.  The results of the
 comparison are reported in
 Table~\ref{pdfstudy}. Figure~\ref{mcfm_pdfx_to_cteq5l} gives the
 ratio of $\sigma$($W^{+}$+2jets for PDF x) to $\sigma$($W^{+}$+2 jets for
 CTEQ5L), while Figure~\ref{mcfm_pdfstudy_ratiowz} shows the ratio of
 $\sigma$($W^{+}$+2 jets) to $\sigma$(Z+2 jets) for a given PDF.  For the
 four PDF's we chose, the changes in the W and Z cross sections
 themselves range from +27\% to -7\% for the W's and +25\% to -8\%
 for the Z's, while the the range of the change in the ratio is from
 +1.5\% to zero, a factor of $\sim 20$ smaller.

\begin{table*}
\caption{\label{pdfstudy}MCFM predictions for the ratios $W^{+}$+2jets and
\Z+2jets with different PDF's.  The PDF's that were compared are
CTEQ3L, CTEQ4L, CTEQ5L, MRSG95, MRST98.  Column one gives the ratio of
$\sigma$($W^{+}$+2jet) at one of the PDF's to $\sigma$($W^{+}$+2jets)
at CTEQ5L.  Column two is the analogous \Z information.  The third
column is the ratio of $R^{+}(2)$ at a specific PDF to $R^{+}(2)$ at CTEQ5L,
where $R^{+}(2)$=~\WpTWOtoZTWO.}
%\begin{tabular}{|l||D{.}{.}{1.3}@{~$\pm$~}D{.}{.}{0.3}|D{.}{.}{1.3}@{~$\pm$~}D{.}{.}{0.3}|D{.}{.}{1.3}@{~$\pm$~}D{.}{.}{0.3}|}
\begin{tabular}{|l||r@{~$\pm$~}l|r@{~$\pm$~}l|r@{~$\pm$~}l|}
\hline
PDF X & \multicolumn{2}{|c|}{($\sigma\WX$) / ($\sigma\WL$)} & 
\multicolumn{2}{|c|}{($\sigma\ZX$) / ($\sigma\ZL$)} &
\multicolumn{2}{|c|}{($\RX$) / ($\RL$)}  \\
\hline\hline
CTEQ5L & ~~1.000 & .000 & ~1.000 & .000 & 1.000 & .000  \\
CTEQ3L & 1.103 & .002 & 1.090 & .002 & 1.011 & .003  \\
CTEQ4L & 1.105 & .002 & 1.094 & .002 & 1.009 & .003  \\
MRSG95 & 1.268 & .002 & 1.249 & .002 & 1.015 & .003  \\
MRST98 & 0.932 & .001 & 0.922 & .001 & 1.011 & .002  \\
\hline
\end{tabular}
\end{table*}

%
% mcfm pdfstudy plot (other pdf)/cteq5l ratio
%

%
% mcfm pdfstudy plot w/z ratio
%

%\subsection{\label{subsection:nlo}Next-To-Leading Order Versus Leading
%Order}

\section{\label{section:sensitivity}Sensitivity to New Contributions to the \WNtoZN ratio} 

A non-Standard Model source of W+ jets or Z + jets would result in a
measured deviation from the expected SM value of the $R_N =\WNtoZN$.
Assuming that the contribution is to W+jets, we can (crudely) estimate
the sensitivity to new physics in each of the $W+$ N-jet channels by
multiplying the uncertainty on the ratio $\WNtoZN$ by the exclusive
$W+N$ jets cross section (if instead the source feeds Z+ jets at the
same crossection, the sensitivity will be larger by a factor of about
10~\cite{maliciousness}.

The estimates above of the systematic uncertainties on $R_N$ are on
the order of several percent; an estimate based on the Run I CDF
experience in measuring R is that 1\% in that ratio 
may be achievable~\cite{Sacha_thesis}.
Statistical uncertainties would then be expected to dominate over
systematics in Run II at the Tevatron for N greater than 2.

Making the assumptions that the new contributions are to the W
cross section and not that of the Z, that the systematics on the ratio
can be reduced with a much larger dataset~\cite{larger_dataset} 
from several percent to 1\%, and that one uses only the electron modes of W
and Z decays, we find the 1-sigma cross-section uncertainties on
new physics shown in Table~\ref{reach}. The muon channel would be
expected to double the statistics (and hence lower the uncertainties
by $\sqrt{2}$). 

\begin{table}[!t]
%\begin{table}[!htb]
\caption{\label{reach}The cross section corresponding to a 1-sigma
uncertainty in the $W/Z$ ratio in 2 fb$^{-1}$, and in 15
fb$^{-1}$. The bins up through N$=$4 use the cross sections of
~\cite{CDF_Wjets}; the N$=$5 and higher bins have been extrapolated
using an exponential, with a factor of 4.8 for each successive
jet. Note that the number of $\Z \rightarrow e^+e^-$ events in each
bin will be approximately a factor of 10 smaller than the
corresponding number of W events. Using the dimuon channel one can
gain a factor of approximately $\sqrt 2$ on these uncertainties.}

\begin{tabular}{|c|r||r|r|}
\hline
\multicolumn{2}{|c|}{Event and W Properties} & \multicolumn{2}{|c|}{W/Z
Ratio Method Reach}\\
\hline
N(Jets) & $\sigma_W$~~~~  & $\sigma_{new}$ 2 $fb^{-1}$ & $\sigma_{new}$
15 $fb^{-1}$ \\
\hline
0 & 1896 pb~~ & 20 pb  (1.0\%)~~ &  20 pb (1.0\%)~~  \\
1 &  370 pb~~ & 4.4 pb (1.2\%)~~ & 3.7 pb (1.0\%)~~  \\
2 &   83 pb~~ & 1.5 pb (1.8\%)~~ & 0.9 pb (1.1\%)~~  \\
3 &   15 pb~~ & 0.5 pb (3.5\%)~~ & 240 fb (1.6\%)~~  \\
4 &  3.1 pb~~ & 230 fb (7.5\%)~~ &  95 fb (2.9\%)~~  \\
5 &  650 fb~~ & 100 fb (16\%)~~  &  40 fb (6\%)~~  \\
6 &  140 fb~~ & 50 fb  (36\%)~~  &  18 fb (13\%)~~  \\
7 &   28 fb~~ & 20 fb  (78\%)~~  &   8 fb (29\%)~~  \\
8 &    6 fb~~ & ------~~~~       &   4 fb (63\%)~~  \\
\hline
\end{tabular}
\end{table}

 Additional sensitivity can come from comparing observed with expected
kinematic distributions or by looking for additional objects in the
events. In particular, the production of a pair of b-quarks suppresses
the cross section over that for light quark production by a large
factor, in principle allowing a corresponding increase in sensitivity.  
Table~\ref{0b_to_2b} shows the ratio of the QCD cross section for
producing N jets including no b quarks to N jets including two b
quarks, for W or Z production. However standard model top production
will provide a large background for non-standard model physics in
these signatures.

%
%W+Njet(0b):W+Njet(2b) and W+Njet(2b):W+Njet(4b) ratios
%
\begin{table*}[!htbp]
\centering
\begin{tabular}{|r@{/}l@{:}c||r@{/}l@{:}c||r@{/}l@{:}c|}
%\multicolumn{6}{|c|}{(N jets)/($\bbbar$+(N-2) jets) for W
%{\it or} Z } \\
\hline
\multicolumn{6}{|c|}{(N jets) / ($\bbbar$+(N-2) jets) for \Wp or \Zg} &
\multicolumn{3}{|c|}{\WpbbNjets/\ZbbNjets} \\
\hline

\Wp+2j & \Wp$\bbbar$+0j & ~90.29 $\pm$ .96 & Z+2j & Z$\bbbar$+0j & ~58.84
$\pm$ .89 & \Wp$\bbbar$+0j & Z$\bbbar$+0j & 1.53 $\pm$ .03 \\

\Wp+3j & \Wp$\bbbar$+1j & 54.72 $\pm$ .84 & Z+3j & Z$\bbbar$+1j &
33.94 $\pm$ .69 & \Wp$\bbbar$+1j & Z$\bbbar$+1j & 1.61 $\pm$ .04 \\

\Wp+4j & \Wp$\bbbar$+2j & 37.58 $\pm$ 1.30 & Z+4j & Z$\bbbar$+2j &
22.83 $\pm$ .40 & \Wp$\bbbar$+2j & Z$\bbbar$+2j & 1.65 $\pm$ .06 \\
%W+5j & W$\bbbar$+3j & & Z+5j & Z$\bbbar$+3j & \\
\hline
\end{tabular}
\caption{Ratios of the cross sections for \WpNjets (including no b
quarks) to \WpNjets (including two b quarks), and ratios of \ZNjets
(including no b quarks) to \ZNjets (including two b quarks).  Also
given are the ratios \WpbbNjets to \ZbbNjets.}
\label{0b_to_2b}
\end{table*}

\section{\label{section:conclusions}Conclusions}

The measurement of the production cross sections of the vector bosons
$W^{\pm}$ and $\Z$ in association with a number (N) of jets is now a
standard way of looking for the production of new particles or
processes that are not described by the Standard Model. With the
expected increased luminosities of Run II and the LHC, N can be quite
large; processes such as associated production of a Higgs boson with a
$\ttbar$ pair can produce W+ 6 jets (4 of which are b-quarks), for
instance. Increasing the precision of the comparison with Standard
Model predictions is necessary, as there are truly difficult problems,
both theoretical and experimental, in predicting the cross sections
for $\WNjets$ and $\ZNjets$ when N is large.

Using the Monte Carlo generators MadGraph and
MCFM at the parton level, and the published CDF data on $W$
and $Z$ + jets production, we have made initial
estimates of the systematic limits on the precision that can be
achieved in the measurement of the ratios of W to Z production,
\WNtoZN, as a function of the number of observed jets, N. The results
indicate that the ratios are at least an order-of-magnitude less
sensitive to experimental and statistical uncertainties than the
individual cross sections.  In particular the ratios are more robust
for large values of N, where the experimental uncertainties in the
energy scale and contributions from the underlying event and multiple
interactions lead to a rapid growth in the cross section uncertainty
with N. 

With respect to the theoretical uncertainties, at N=2, for example, we
find the uncertainty due to choice of \Qsq~scale is a factor of $\sim
8$ smaller in the ratio \WNtoZN than in the individual W or Z cross
sections. Similarly, the uncertainty due to the choice of PDF, largely
driven by the $u/d$ quark ratio, is smaller in the ratio by a factor
of $\sim$20.

The experimental uncertainties in the cross sections, dominated by the
uncertainty in the jet energy scale and contributions from the
underlying event, are greatly diminished by focusing on the ratio of
W and \Zg cross sections ratio than the cross sections themselves.  In
particular the uncertainty due to uncertainties in the jet energy
scale, the contributions from the underlying event, multiple
interactions in one event, etc. cancel to a high degree. We have here made
estimates at the parton level;  a
full determination of these will require the new data and a full
analysis; our initial estimates are that the ratios can be determined
at the several percent level. This is a significant improvement over
the present uncertainties on the cross sections themselves.

%\clearpage

\begin{acknowledgments}
It is a pleasure to acknowledge helpful conversations with Edward
Boos, John Campbell, Jay Dittman, Lev Dudko, Keith Ellis, Michelangelo
Mangano, Stephen Mrenna, Jon Rosner, and Tim Stelzer. Special thanks
are due to Tim Stelzer and John Campbell for help with MadGraph and
MCFM, respectively.  This work was supported by the National Science
Foundation, grant PHY02-01792.
\end{acknowledgments}

%\clearpage
% Be careful- this has to be changed if you change the bib file name!
\bibliography{WZratio_arXiv_version}

\begin{thebibliography}{60}
\bibitem[{plu()}]{plus_only}
\bibinfo{note}{Because the Monte Carlo programs specifically ask for the sign
  of the final state charged lepton, we use the production crossection for only
  one sign ($W^+$) of W production, resulting in ratios $R^+$ which are half
  ($\sim 5$) of the usually quoted $R = (W^+ + W^-)/\Zg$ ($\sim 10$).}
\bibitem[{Zga()}]{Zgamma_poles}
\bibinfo{note}{The production in the charged dilepton channels (e.g. $\epem$)
  proceed through two s-channel amplitudes, the $\gamma$ and $\Z$ poles. While
  the $\Z$ pole dominates for the selection criteria used here ($\sim 96\%$),
  both amplitudes must always be included (experimentalists have evolved the
  jargon that $\Z$ means both amplitudes, as the source of the resulting final
  state that is measured by nature cannot be untangled.)}.
\bibitem[{\citenamefont{Spiropulu}(2000.)}]{Maria}
\bibinfo{author}{\bibfnamefont{M.}~\bibnamefont{Spiropulu}}, Ph.D. thesis,
  \bibinfo{school}{Harvard University} (\bibinfo{year}{2000.}),
  \bibinfo{note}{this is the first use of the ratios \WNtoZN we have found;
  here they are used to normalize the W + N jets prediction to the Z + 2 jet
  data.}
\bibitem[{EtP()}]{EtPt}
\bibinfo{note}{The transverse momentum is defined as $\Pt = p\sin\theta$; the
  transverse energy is defined as $\Et = E\sin\theta$. Missing transverse
  energy is defined as $\mett = -\Sigma\Et$, where the sum is over all objects
  in an event. We use the convention that `momentum' refers to $pc$ and `mass'
  to $mc^2$, so that energy, momentum, and mass are all measured in GeV.}
\bibitem[{inc()}]{inclusive}
\bibinfo{note}{By `inclusive' we mean counting the number of jets $\ge N$ for
  each N; by `exclusive' we mean counting exactly N jets.}
\bibitem[{mul()}]{multiple_interactions}
\bibinfo{note}{The proton and antiproton beams each consist of 36 bunches
  distributed around the 1-km radius Tevatron ring. Proton-antiproton
  interactions occur when the two beam cross each other; more than one
  collision can occur in a single crossing. The time-response of the detectors
  is typically such that the effects of multiple interactions are integrated
  over, with only partial discrimination between interactions being possible by
  reconstructing charged tracks to different vertex positions.}
\bibitem[{apo()}]{apologia}
\bibinfo{note}{The measurement and understanding of the jet $E_T$ and $\eta$
  distributions is a major task for Run II and the LHC; the focus here is
  instead on the extent that the W and \Zg cases cancel, and hence an estimate
  of the expected precision on the ratio.}
\bibitem[{jet()}]{jet_balancing}
\bibinfo{note}{As the calorimeter response is calibrated by jet balancing in
  $\eta$ the energy scales have a maximum excursion versus $\eta$ of a few \%.
  It is this uncertainty times the difference in the jet $\eta$ distributions
  that sets the limit on the contributions to the ratio from the jet energy
  scale.}
\bibitem[{mal()}]{maliciousness}
\bibinfo{note}{Of course it is possible that a new source feeds both numerator
  and denominator in equal proportion; in that case this method has no
  sensitivity.}
\bibitem[{lar()}]{larger_dataset}
\bibinfo{note}{With the larger dataset the uncertainties on such quantities as
  the u/d ratio, and hence the PDFs, the $\eta$ and $\Pt$ distributions of W
  and Z production, and other quantities will also be better measured}.
\end{thebibliography}
% WZratio_preprint.bib

\end{document}